\documentclass{article}
\usepackage{ijcai19}

\usepackage{times}
\usepackage{soul}
\usepackage{url}
\usepackage[draft]{hyperref}
\usepackage[utf8]{inputenc}
\usepackage[small]{caption}
\usepackage{graphicx}
\usepackage{amsmath}
\usepackage{booktabs}
\usepackage{algorithm}
\usepackage{algorithmic}
\usepackage{bibentry}
\usepackage{amsfonts}
\usepackage{subfigure}
\usepackage{multirow}
\usepackage{epsfig}
\usepackage{color}
\usepackage{epstopdf}
\usepackage{rotating}
\usepackage{caption}
\usepackage{multicol}
\usepackage{amssymb}
\usepackage{bbding}
\usepackage{float}
\usepackage{mathtools}
\usepackage[medium,compact]{titlesec}

\title{Deep Adversarial Social Recommendation}

\author{
Wenqi Fan$^1$\and
Tyler Derr$^2$\and
Yao Ma$^{2}$\and
Jianping Wang$^1$\and
Jiliang Tang$^2$\And
Qing Li$^3$
\affiliations
$^1$Department of Computer Science, City University of Hong Kong\\
$^2$Data Science and Engineering Lab, Michigan State University\\
$^3$Department of Computing,The Hong Kong Polytechnic University
\emails
wenqifan03@gmail.com,
\{derrtyle, mayao4\}@msu.edu,
jianwang@cityu.edu.hk,
tangjili@msu.edu,
csqli@comp.polyu.edu.hk
}

\begin{document}

\maketitle

\begin{abstract}
Recent years have witnessed rapid developments on social recommendation techniques for improving the performance of recommender systems due to the growing influence of social networks to our daily life. The majority of existing social recommendation methods unify user representation for the user-item interactions (item domain) and user-user connections (social domain). However, it may restrain user representation learning in each respective domain, since users behave and interact differently in the two domains, which makes their representations to be heterogeneous. In addition, most of traditional recommender systems can not efficiently optimize these objectives, since they utilize negative sampling technique which is unable to provide enough informative guidance towards the training during the optimization process. In this paper, to address the aforementioned challenges, we propose a novel deep adversarial social recommendation framework \textbf{DASO}. It adopts a bidirectional mapping method to transfer users' information between social domain and item domain using adversarial learning. Comprehensive experiments on two real-world datasets show the effectiveness of the proposed framework.
\end{abstract}

\section{Introduction}


In recent years, we have seen an increasing amount of attention on social recommendation, which harnesses social relations to boost the performance of recommender systems~\cite{tang2016recommendation,fan2019Graph,wang2016social}. Social recommendation is based on the intuitive ideas that people in the same social group are likely to have similar preferences, and that users will gather information from their experienced friends (e.g., classmates, relatives, and colleagues) when making decisions. Therefore, utilizing users' social relations has been proven to greatly enhance the performance of many recommender systems~\cite{ma2008sorec,fan2019Graph,tang2013social,tang2016recommendations}.

\begin{figure}[tbp]
\centering
\vskip -0.20in
{\includegraphics[width=0.75\linewidth]{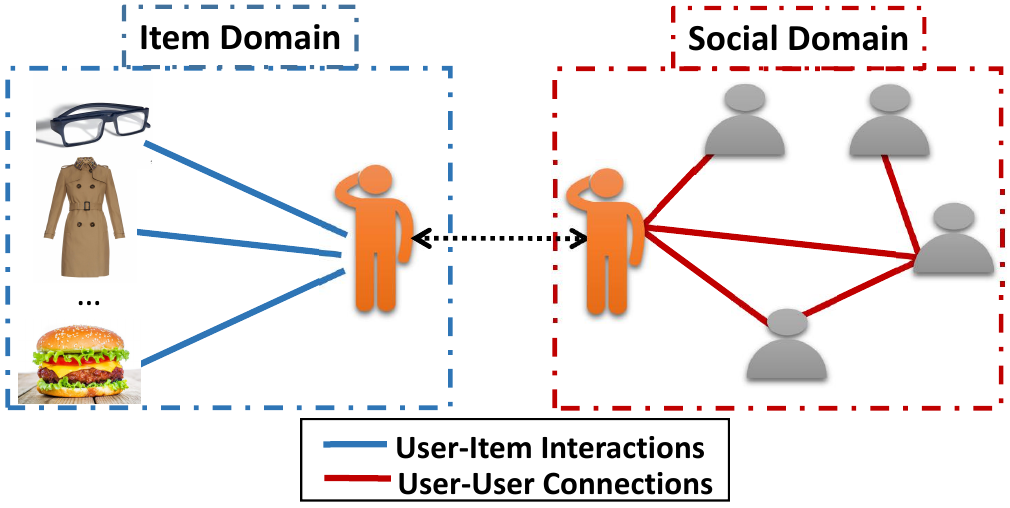}}
\vskip -1ex
\caption{An illustration of one user in two domains (Item Domain and Social Domain) for social recommendations.}\label{fig:intro}
\vskip -0.20in
\end{figure}

In Figure~\ref{fig:intro}, we observe that in social recommendation we have both the item and social domains, which represent the user-item interactions and user-user connections, respectively. Currently, the most effective way to incorporate the social relation information for improving recommendations is when learning user representations, which is commonly achieved in ways such as, using trust propagation~\cite{jamali2010matrix}, incorporating a user's social neighborhood information~\cite{fan2018deep}, or sharing a common user representation for the user-item interactions and social relations with a co-factorization method~\cite{ma2008sorec}. However, as shown in Figure~\ref{fig:intro}, although users bridge the gap between these two domains, their representations should be heterogeneous. This is because users behave and interact differently in the two domains. Thus, using a unified user representation may restrain user representation learning in each respective domain and results in an inflexible/limited transferring of knowledge from the social relations to the item domain. Therefore, one challenge is to learn separated user representations in the two domains while transferring the information from the social domain to the item domain for social recommendation.

In this paper, we adopt a nonlinear mapping operation to transfer user's information from the social domain to the item domain, while learning separated user representations in the two domains. Nevertheless, learning the representations is challenging due to the inherent data sparsity problem in both domains. Thus, to alleviate this problem, we propose to use a bidirectional mapping between the two domains, such that we can cycle information between them to progressively enhance the user's representations in both domains. However, for optimizing the user representations and item representations, most existing methods utilize the negative sampling technique, which is quite ineffective~\cite{wang2018neural}. This is due to the fact that during the beginning of the training process, most of the negative user-item samples are still within the margin to the real user-item samples, but later during the optimization process, negative sampling is unable to provide ``difficult'' and informative samples to further improve the user representations and item representations~\cite{wang2018neural,cai2018kbgan}. Thus, it is desired to have samples dynamically generated throughout the training process to better guide the learning of the user representations and item representations.

Recently, Generative Adversarial Networks (GANs)~\cite{goodfellow2014generative,Derr2019DeepAN}, which consists of two models to process adversarial learning, have shown great success across various domains due to their ability to learn an underlying data distribution and generate synthetic samples~\cite{mao2017least,mao2018effectiveness,brock2019large,liu2018generative,wang2017irgan,wang2018graphgan}. This is performed through the use of a generator and a discriminator. The generator tries to generate realistic fake data samples to fool the discriminator, which distinguishes whether a given data sample is produced by the generator or comes from the real data distribution. A minimax game is played between the generator and discriminator, where this adversarial learning can train these two models simultaneously for mutual promotion. In~\cite{wang2018neural} adversarial learning had been used to address the limitation of typical negative sampling. Thus, we propose to harness adversarial learning in social recommendation to generate ``difficult'' negative samples to guide our framework in learning better user and item representations while further utilizing it to optimize our entire framework.

In this paper, we propose a \textbf{D}eep \textbf{A}dversarial \textbf{SO}cial recommendation framework \textbf{DASO}. Our major contributions can be summarized as follows:
\begin{list}{\labelitemi}{\leftmargin=1em}
    \setlength{\topmargin}{0pt}
    \setlength{\itemsep}{0em}
    \setlength{\parskip}{0pt}
    \setlength{\parsep}{0pt}
\item We introduce a principled way to transfer users' information from social domain to item domain using a bidirectional mapping method where we cycle information between the two domains to progressively enhance the user representations; 
\item We propose a deep adversarial social recommender system DASO, which can harness the power of adversarial learning to  dynamically generate ``difficult'' negative samples, learn the bidirectional mappings between the two domains, and ultimately optimize better user and item representations; and 
\item We conduct comprehensive experiments on two real-world datasets to show the effectiveness of the proposed framework.
  \end{list}


\begin{figure*}[tbp]
\centering
\vskip -2ex
{\includegraphics[width=0.85\linewidth]{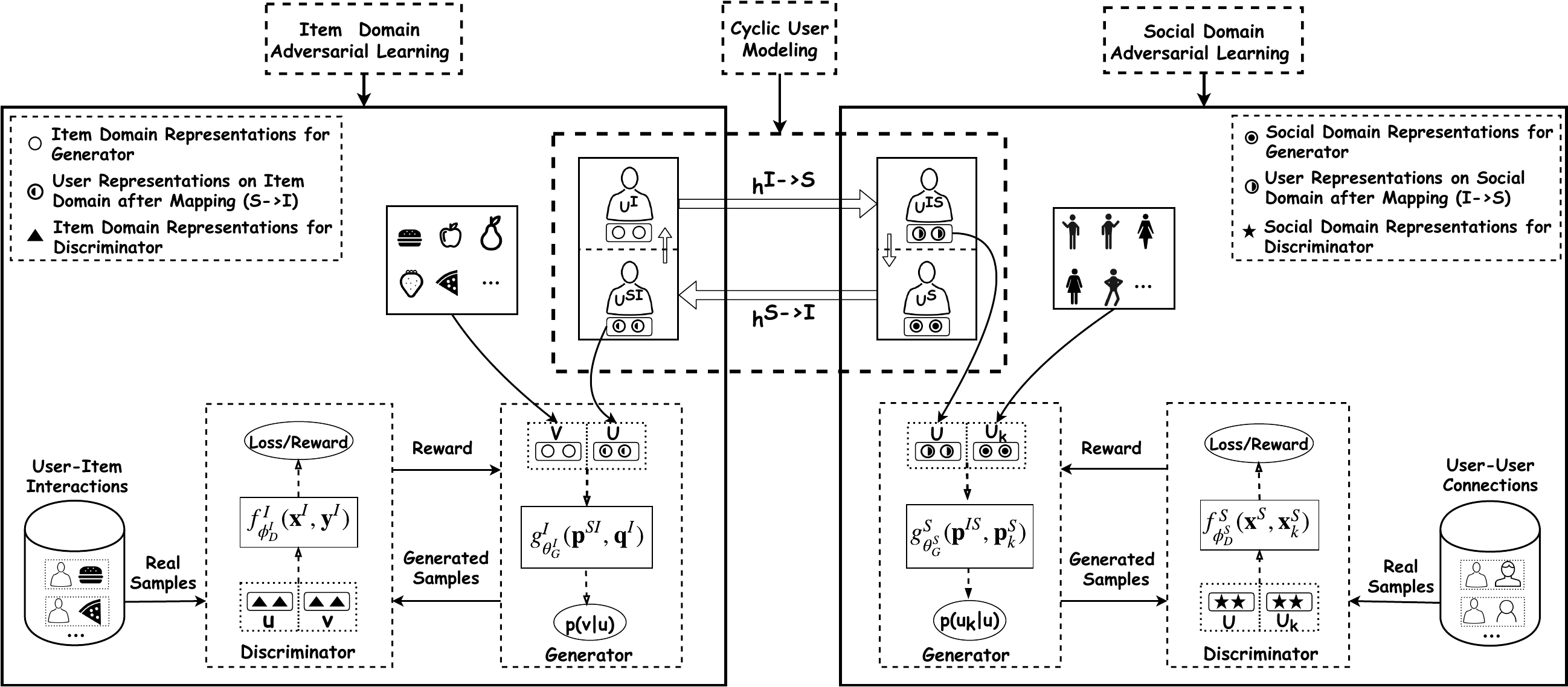}}
\caption{The overall architecture of the proposed model. \label{fig:framework}}
\vskip -0.20in
\end{figure*}

\section{The Proposed Framework}
\label{sec:framework}

We first introduce definitions and notations that are used through the paper. Let $ \mathcal{U}= \left \{ u_1, u_2, ..., u_N \right \}$ and $ \mathcal{V} = \left \{ v_1, v_2, ..., v_M \right \}$ denote the sets of users and items respectively, where $N$ is the number of users, and $M$ is the number of items. We define user-item interactions matrix ${\bf R} \in  \mathbb{R}^{N\times M}$ from user's implicit feedback, where the $i,j$-th element $r_{i,j}$ is 1 if  there is an interaction (e.g., clicked/bought) between user $u_i$ and item $v_j$, and $0$ otherwise. However, $r_{i,j} = 1$ does not mean user $u_i$ actually likes item $v_j$. Similarly, $r_{i,j} = 0$ does not mean $u_i$ does not like item $v_j$, since it can be that the user $u_i$ is not aware of the item $v_j$. The social network between users can be described by a matrix $ {\bf S}\in \mathbb{R}^{N\times N}$, where $s_{i,j}=1$ if there is a social relation between user $u_i$ and user $u_j$, and $0$ otherwise. Given the interaction matrix ${\bf R}$ and the social network ${\bf S}$, we aim to predict the unobserved entries (i.e., those where $r_{i,j} = 0$) in ${\bf R}$.


\subsection{An Overview of the Proposed Framework}

The architecture of the proposed model is shown in Figure~\ref{fig:framework}. The information is from two domains, which are the item domain $I$ and the social domain $S$. The model consists of three components: cyclic user modeling, item domain adversarial learning, and social domain adversarial learning. The cyclic user modeling is to model user representations on two domains.
The item domain adversarial learning is to adopt the adversarial learning for dynamically generating ``difficult'' and informative negative samples  to guide the learning of user and item representations. The generator is utilized to `sample' (recommend) items for each user and output user-item pairs as fake samples; the other is the discriminator, which distinguishes the user-item pair samples sampled from the real user-item interactions from the generated user-item pair samples. The social domain adversarial learning also similarly consists of a generator and a discriminator. 

There are four types of representations in the two domains. In the item domain $I$, we have two types of representations including item domain representations of the generator  (${\bf p}_{i}^{I} \in \mathbb{R}^d$  for user $u_i$ and ${\bf q}_{j}^{I} \in \mathbb{R}^d$ for item $v_j$), and the item domain representations of the discriminator (${\bf x}_{i}^{I} \in \mathbb{R}^d$ for user $u_i$ and ${\bf y}_{j}^{I} \in \mathbb{R}^d$ for item $v_j$). Social domains $S$ also contains two types of representations including the social domain representations of the generator (${\bf p}_{i}^{S} \in \mathbb{R}^d$  for user $u_i$), and the social domain representations of the discriminator (${\bf x}_{i}^{S} \in \mathbb{R}^d$  for user $u_i$). Next we discuss the details for each component.

%



\subsection{Cyclic User Modeling}

Cyclic user modeling aims to learn a relation between the user representations in the item domain $I$ and the social domain $S$. As shown in the top part of Figure~\ref{fig:framework},  we first adopt a nonlinear mapping operation, denoted as $h^{S \rightarrow I}$, to transfer user's information from the social domain to the item domain, while learning separated user representations in the two domains. Then, a bidirectional mapping between these two domains (achieved by including another nonlinear mapping $h^{I \rightarrow S}$) is utilized to help cycle the information between them to progressively enhance the user representations in both domains.



\subsubsection{Transferring Social Information to Item Domain}

In social networks, a person's preferences can be influenced by their social interactions, suggested by sociologists~\cite{fan2019Graph,fan2018deep,wasserman1994social}. Therefore, a user's social relations from the social network should be incorporated into their user representation in the item domain.

We propose to adopt nonlinear mapping operation to transfer user's information from the social domain to the item domain. More specifically, the user representation on social domain ${\bf p}_{i}^{S}$ is transferred to the item domain via a Multi-Layer Perceptron (MLP) denoted as $h^{S \rightarrow I}$. The transferred user representation from social domain is denoted as ${\bf p}_{i}^{SI}$. More formally, the nonlinear mapping is as follows: $    {\bf p}_{i}^{SI} = h^{S \rightarrow I}({\bf p}_{i}^{S}) = {\bf W}_L\cdot(\cdots a( {\bf W}_2\cdot a({\bf W}_1\cdot {\bf p}_{i}^{S} +{\bf b}_1 )+ {\bf b}_2 ) \dots) + {\bf b}_L$,
where the ${\bf W}$s, ${\bf b}$s are the weights and biases for the layers of the neural network having $L$ layers, and $a$ is a non-linear activation function.


\subsubsection{Bidirectional Mapping with Cycle Reconstruction}

As user-item interactions and user-user connections are often very sparse, learning separated user representations is challenging. Therefore, to partially alleviate this issue, we propose to utilize a bidirectional mapping between the two domains, such that we can cycle information between them to progressively enhance the user representations in both domains. To achieve this, another nonlinear mapping operation, denoted as $h^{I \rightarrow S}$, is adopted to transfer information from the item domain to the social domain: ${\bf p}_{i}^{IS} = h^{I \rightarrow S}({\bf p}_{i}^{I})$, which has the same network structure as the $h^{S \rightarrow I}$.


This Bidirectional Mapping allows knowledge to be transferred between item and social domains. To learn these mappings, we further introduce cycle reconstruction. Its intuition is that transferred knowledge in the target domain should be reconstructed to the original knowledge in the source domain.  Next we will elaborate cycle reconstruction.

For user  $u_i$'s item domain representation ${\bf p}_{i}^{I}$, the user representation with cycle reconstruction should be able to map ${\bf p}_{i}^{I}$ back to the original domain, as follows, ${\bf p}_{i}^{I} \rightarrow h^{I \rightarrow S}({\bf p}_{i}^{I}) \rightarrow  h^{S \rightarrow I}(h^{I \rightarrow S}({\bf p}_{i}^{I})) \approx {\bf p}_{i}^{I}$.
Likewise, for user $u_i$'s social domain representation ${\bf p}_{i}^{S}$,  the user representation with cycle reconstruction can also bring ${\bf p}_{i}^{S}$ back to the original domain: ${\bf p}_{i}^{S} \rightarrow h^{S \rightarrow I}({\bf p}_{i}^{I}) \rightarrow  h^{I \rightarrow S}(h^{S \rightarrow I}({\bf p}_{i}^{S})) \approx {\bf p}_{i}^{S}$.

We can formulate this procedure using a cycle reconstruction loss, which needs to be minimized, as follows,
\begin{small}
\vspace*{-0.5\baselineskip}
\begin{align}
\mathcal{L}_{cyc} (h^{S \rightarrow I}, h^{I \rightarrow S}) \label{eq:cyc}  & = \sum_{i=1}^{N}\left (   \left \|  h^{S \rightarrow I}(h^{I \rightarrow S}({\bf p}_{i}^{I}))  - {\bf p}_{i}^{I}  \right \|_2     \right.  \\
   \nonumber & \left.+   \left \|  h^{I \rightarrow S}(h^{S \rightarrow I}({\bf p}_{i}^{S}))  - {\bf p}_{i}^{S}  \right \|_2    \right )
\end{align}
\end{small}



\subsection{Item Domain Adversarial Learning}

To address the limitation of negative sampling for recommendation on the ranking task, we propose to harness adversarial learning to generate ``difficult'' and informative samples to guide the framework in learning better user and item representations in the item domain.  As shown in the bottom left part of Figure~\ref{fig:framework}, the adversarial learning on item domain consists of two components:

\textbf{Discriminator} $D^I (u_i, v;\phi^I_D)$, parameterized by $\phi^I_D$, aims to distinguish the real user-item pairs $(u_i, v)$ and the user-item pairs generated by the generator.

\textbf{Generator} $G^I(v | u_i;\theta^I_G)$, parameterized by $\theta^I_G$, tries to fit the underlying real conditional distribution $p^I_{real}  (v|u_i)$ as much as possible, and generates (or, to be more precise, selects) the most relevant items to a given user $u_i$.

Formally, $D^I$ and $G^I$ are playing the following two-player minimax game with value function $\mathcal{L}_{adv}^I(G^I, D^I)$,
\begin{small}
\begin{align}
 &  min_{\theta^I_G} max_{\phi^I_D} \mathcal{L}_{adv}^I(G^I, D^I) \label{eq:obj}\\
\nonumber  & = \sum_{i=1}^{N}\left ( \mathbb{E}_{v \sim p^I_{real}(\cdot |u_i) }   \left [ log D^{I}(u_i, v;\phi^I_D) \right ]    \right. \\
\nonumber  & \left.  +  \mathbb{E}_{v \sim G^I(\cdot |u_i;\theta^I_G) }   \left  [ log (1 - D^{I}(u_i, v;\phi^I_D)) \right ]    \right )
\end{align}
\end{small}


\subsubsection{Item Domain Discriminator Model}
Discriminator $D^I$ aims to distinguish real user-item pairs (i.e., real samples) and the generated ``fake'' samples. The discriminator $D^I$ estimates the probability of item $v_j$  being relevant (bought or clicked) to a given user $u_i$ using the sigmoid function as follows: $D^I (u_i, v_j;\phi^I_D) = \sigma (f^I_{\phi^I_D}({\bf x}_{i}^{I}, {\bf y}_{j}^I) ) =  \frac{1}{1+exp(- f^I_{\phi^I_D}({\bf x}_{i}^{I}, {\bf y}_{j}^I))} $,
where $f^I_{\phi^I_D}$ is a score function.

Given real samples and generated fake samples, the objective for the discriminator $D^I$ is to maximize the log-likelihood of assigning the correct labels to both real and generated samples. The discriminator can be optimized by minimizing the objective in eq.~\eqref{eq:obj} with the generator fixed using stochastic gradient methods.
\subsubsection{Item Domain Generator Model}


On the other hand, the purpose of the generator $G^I$ is to approximate the underlying real conditional distribution $p^I_{real}  (v|u_i)$, and generate the most relevant items for any given user $u_i$.

We define the generator using the softmax function over all the items according to the transferred user representation ${\bf p}_{i}^{SI}$ from social domain to item domain: $G^I(v_j | u_i;\theta^I_G) = \frac{exp(g^I_{\theta^I_G} ({\bf p}_{i}^{SI}, {{\bf q}_j^I}))}{\sum_{v_j \in \mathcal{V} }exp(g^I_{\theta^I_G}({\bf p}_{i}^{SI}, {{\bf q}_j^I}))}$,
where $g^I_{\theta^I_G}$ is a score function reflecting the chance of $v_j$ being clicked/purchased by $u_i$. Given a user $u_i$, an item $v_j$ can be sampled from the distribution $G^I(v_j | u_i;\theta^I_G)$.

We note that the process of generating a relevant item for a given user is discrete. Thus, we cannot optimize the generator $G^I$ via stochastic gradient descent methods~\cite{wang2017irgan}. Following ~\cite{sutton2000policy,schulman2015gradient}, we adopt the policy gradient method usually adopted in reinforcement learning to optimize the generator.

To learn the parameters for the generator, we need to perform the following minimization problem:
\begin{small}
\begin{align}
 &  min_{\theta^I_G}  \sum_{i=1}^{N}\left(    \mathbb{E}_{v \sim G^I(\cdot |u_i;\theta^I_G) }   \left[ log (1 - D^{I}(u_i, v;\phi^I_D)) \right]\right)
\end{align}
\end{small}
which is equivalent to the following maximization problem
\begin{tiny}
\begin{align}
 &  max_{\theta^I_G}  \sum_{i=1}^{N}\left(    \mathbb{E}_{v \sim G^I(\cdot |u_i;\theta^I_G) }   \left[\log ( 1+exp( f^I_{\phi^I_D}({\bf x}_{i}^{I}, {\bf y}_{j}^I))) \right]\right)
\end{align}
\end{tiny}

Now, this problem can be viewed in a reinforcement learning setting, where $\log(1+exp( f^I_{\phi^I_D}({\bf x}_{i}^{I}, {\bf y}_{j}^I)))$ is the reward given to the action ``selecting $v_i$ given a user $u_i$'' performed according to the policy probability $G^I(v |u_i)$. The policy gradient can be written as:

\begin{tiny}
\begin{align}\label{eq:policygradient}
& \bigtriangledown_{\theta^I_G}   \mathcal{L}_{adv}^I(G^I, D^I) \\
 &=\sum_{i=1}^{N} \sum_{j=1}^{M}  \bigtriangledown_{\theta^I_G} G^I( v_j |u_i) \ \log( 1+exp( f^I_{\phi^I_D}({\bf x}_{i}^{I}, {\bf y}_{j}^I))) \\
 &= \sum_{i=1}^{N} \sum_{j=1}^{M}  G^I( v_j |u_i) \bigtriangledown_{\theta^I_G} \ log G^I( v_j |u_i) \ \log(1+exp( f^I_{\phi^I_D}({\bf x}_{i}^{I}, {\bf y}_{j}^I))) \\
 &= \sum_{i=1}^{N}  \mathbb{E}_{v_j \sim G^I(\cdot |u_i) } \left  [ \bigtriangledown_{\theta^I_G} \ log G^I( v_j |u_i) \ \log( 1+exp( f^I_{\phi^I_D}({\bf x}_{i}^{I}, {\bf y}_{j}^I))) \right ]
\end{align}
\end{tiny}

Specially, the gradient $\bigtriangledown_{\theta^I_G}   \mathcal{L}_{adv}^I(G^I, D^I)$ is an expected summation over the gradients $\bigtriangledown_{\theta^I_G}  log G^I(v_j | u_i)$ weighted by $ \log(1+exp( f^I_{\phi^I_D}({\bf x}_{i}^{I}, {\bf y}_{j}^I)))$.


The optimal parameters of $G^I$ and $D^I$ can be learned by alternately minimizing and maximizing the value function $\mathcal{L}^I_{adv}(G^I, D^I)$. In each iteration, discriminator $D^I$ is trained with real samples from  $p^I_{real}  (\cdot|u_i)$ and generated samples from generator $G^I$; the generator $G^I$ is updated with policy gradient under the guidance of $D^I$.

Note that different from the way of optimizing user and item representations with the typical negative sampling on traditional recommender systems, the adversarial learning technique tries to generate ``difficult'' and high-quality negative samples to guide the learning of user and item representations.

\subsection{Social Domain Adversarial Learning}

In order to learn better user representations from the social perspective, another adversarial learning is harnessed in the social domain. Likewise, the adversarial learning in the social domain consists of two components, as shown in the bottom right part of Figure~\ref{fig:framework}.

\textbf{Discriminator} $D^S (u_i,u;\phi^S_D)$, parameterized by $\phi^S_D$,   aims to distinguish the real connected user-user pairs $(u_i, u)$ and the fake user-user pairs generated by the generator $G^S$.

\textbf{Generator} $G^S(u | u_i;\theta^S_G)$, parameterized by $\theta^S_G$, tries to fit the underlying real conditional distribution $p^S_{real}  (u|u_i)$ as much as possible, and generates (or, to be more precise, selects) the most relevant users to the given user $u_i$.

Formally, $D^S$ and $G^S$ are playing the following two-player minimax game with value function $\mathcal{L}_{adv}^S(G^S, D^S)$,
\begin{small}
\begin{align}
 &  min_{\theta^S_G} max_{\phi^S_D} \mathcal{L}_{adv}^S(G^S, D^S) \\
\nonumber  & = \sum_{i=1}^{N}\left ( \mathbb{E}_{u \sim p^S_{real}(\cdot |u_i) }   \left [ log D^{S}(u_i, u;\phi^S_D) \right ]    \right. \\
\nonumber  & \left.  +  \mathbb{E}_{u \sim G^S(\cdot |u_i;\theta^S_G) }   \left  [ log (1 - D^{S}(u_i, u;\phi^S_D)) \right ]    \right )
\end{align}
\end{small}

\subsubsection{Social Domain Discriminator}
The discriminator $D^S$ aims to distinguish the real user-user pairs and the generated ones. The discriminators $D^S$ estimates the probability of user $u_k$  being connected to user $u_i$ with a sigmoid function as follows: $D^S (u_i, u_k;\phi^S_D)  = \sigma (f^S_{\phi^S_D}({\bf x}_{i}^{S}, {\bf x}_{k}^S) ) =  \frac{1}{1+exp(- f^S_{\phi^S_D}({\bf x}_{i}^{S}, {\bf x}_{k}^S))}$,
where $f^S_{\phi^S_D}$ is a score function.

\subsubsection{Social Domain Generator}
The purpose of the generator, $G^S$, is to approximate the underlying real conditional distribution $p^S_{real}  (u|u_i)$, and generate (or, to be more precise, select) the most relevant users for any given user $u_i$.

We model the distribution using a softmax function over all the other users with the transferred user representation ${\bf p}_{i}^{IS}$ (from the item to social domain),
\begin{small}
\begin{align}
G^S(u_k | u_i;\theta^S_G) = \frac{exp(g^S_{\theta^S_G} ({\bf p}_{i}^{IS}, {{\bf p}_k^S}))}{\sum_{u_k \neq u_i }exp(g^S_{\theta^{S}_G}({\bf p}_{i}^{IS}, {{\bf p}_k^S}))}
\end{align}
\end{small}
where $g^S_{\theta^S_G}$ is a score function reflecting the chance of $u_k$ being related to $u_i$.

Likewise, policy gradient is utilized to optimize the generator $G^S$,
\begin{tiny}
 \begin{align}
&\bigtriangledown_{\theta^S_G}   \mathcal{L}_{adv}^S(G^S, D^S) \\
        &= \sum_{i=1}^{N}  \mathbb{E}_{u_k \sim G^S(\cdot |u_i) } \left  [ \bigtriangledown_{\theta^S_G} \ log G^S( u_k |u_i) \ \log(1+exp(f^S_{\phi^S_D}({\bf x}_{i}^{S}, {\bf x}_{k}^S))) \right ] \label{eq:abc}
\end{align}
\end{tiny}
where the details are omitted here, since it is defined similar to Eq.(\ref{eq:policygradient}).

\subsection{The Objective Function}

With all model components, the objective function of the proposed framework is:
\begin{small}
\begin{align}
&min_{G^I,G^S,h^{S \rightarrow I}, h^{I \rightarrow S} } max_{ D^I, D^S} \mathcal{L} \\ \nonumber
&= F(G^I, D^I, G^S,D^S, h^{S \rightarrow I}, h^{I \rightarrow S}) \\ \nonumber
&= \mathcal{L}^I_{adv}(G^I,D^I)   + \mathcal{L}^S_{adv}(G^S,D^S) + \lambda \mathcal{L}_{cyc}( h^{S \rightarrow I}, h^{I \rightarrow S})
\end{align}
\end{small}
where $\lambda$ is to control the relative importance of cycle-reconstruction strategy and further influences the two mapping operation. $h^{S \rightarrow I}$ and $h^{I \rightarrow S}$ are implemented as MLP with three hidden layers. To optimize the objective, the RMSprop~\cite{Tieleman2012} is adopted as the optimizer in our implementation.
To train our model, at each training epoch, we iterate over the training set in mini-batch to train each model (e.g., $G^I$) while the parameters of other models (e.g.,$D^I, G^S,D^S$) are fixed. When the training is finished, we take the representations learned by the generator $G^I$ and $G^S$ as our final representations of item and user for performing recommendation.

There are six representations in our model, including ${\bf p}_{i}^{I}, {\bf q}_{j}^{I}, {\bf x}_{i}^{I}, {\bf y}_{j}^{I}, {\bf p}_{i}^{S}, {\bf x}_{i}^{S}$. They are randomly initialized and jointly learned during the training stage.

Following the setting of IRGAN~\cite{wang2017irgan}, we adopt the inner product as the score function $f^I_{\phi^I_D}$ and  $g^I_{\theta^I_G}$ in the item domain as follows: $ f^I_{\phi^I_D}({\bf x}_{i}^{I}, {\bf y}_{j}^I)  =  ({\bf x}_{i}^{I})^T {\bf y}_{j}^I  + a_j,g^I_{\theta^I_G} ({\bf p}_{i}^{SI}, {{\bf q}_j^I})  = ({\bf p}_{i}^{SI})^T {{\bf q}_j^I} + b_j$,
where $a_j$ and $b_j$ are the bias term for item $j$. We define the score function $f^S_{\phi^S_D}$ and  $g^S_{\theta^S_G}$ in the social domain in a similar way.
 Note that the above score functions can be also implemented using deep neural networks, but leave this investigation as one future work.

\section{Experiments}
\label{sec:experiments}


\subsection{Experimental Settings}

\begin{table*}[htbp]
\centering
\vskip -0.2in
\caption{Performance comparison of different recommender systems}
\label{tab:baselines_results}
\vskip -0.05in
\scalebox{0.85}{
\begin{tabular}{|c|c|c|c|c|c|c|c|c|}
\hline
\multirow{2}{*}{Datasets} & \multirow{2}{*}{Metrics} & \multicolumn{7}{c|}{Algorithms}                                   \\ \cline{3-9} 
                          &                          & BPR    & IRGAN  & SBPR   & SocialMF & DeepSoR & GraphRec & \textbf{DASO}   \\ \hline \hline
\multirow{6}{*}{Ciao}     & Precision@3                      & 0.0154 & 0.0274 & 0.0211 & 0.0260   & 0.0310  & 0.0374   & \textbf{0.0462} \\ \cline{2-9} 
                          & Precision@5                      & 0.0137 & 0.0245 & 0.0204 & 0.0218   & 0.0240  & 0.0326   & \textbf{0.0451} \\ \cline{2-9} 
                          & Precision@10                     & 0.0102 & 0.0239 & 0.0178 & 0.0155   & 0.0201  & 0.0265   & \textbf{0.0375} \\ \cline{2-9} 
                          & NDCG@3                   & 0.0254 & 0.0337 & 0.0316 & 0.0312   & 0.0380  & 0.0392   & \textbf{0.0509} \\ \cline{2-9} 
                          & NDCG@5                   & 0.0299 & 0.0350 & 0.0335 & 0.0364   & 0.0356  & 0.0373   & \textbf{0.0514} \\ \cline{2-9} 
                          & NDCG@10                  & 0.0315 & 0.0376 & 0.0379 & 0.0373   & 0.0396  & 0.0382   & \textbf{0.0518} \\ \hline \hline
\multirow{6}{*}{Epinions} & Precision@3                      & 0.0046 & 0.0138 & 0.0096 & 0.0100   & 0.0105  & 0.0156   & \textbf{0.0208} \\ \cline{2-9} 
                          & Precision@5                      & 0.0042 & 0.0104 & 0.0089 & 0.0090   & 0.0098  & 0.0123   & \textbf{0.0173} \\ \cline{2-9} 
                          & Precision@10                     & 0.0035 & 0.0080 & 0.0066 & 0.0071   & 0.0086  & 0.0102   & \textbf{0.0140} \\ \cline{2-9} 
                          & NDCG@3                   & 0.0099 & 0.0175 & 0.0136 & 0.0176   & 0.0160  & 0.0183   & \textbf{0.0226} \\ \cline{2-9} 
                          & NDCG@5                   & 0.0128 & 0.0177 & 0.0152 & 0.0196   & 0.0183  & 0.0182   & \textbf{0.0217} \\ \cline{2-9} 
                          & NDCG@10                  & 0.0169 & 0.0202 & 0.0198 & 0.0202   & 0.0200  & 0.0217   & \textbf{0.0234} \\ \hline
\end{tabular}}
\vskip -0.10in
\end{table*}

We conduct our experiments on two representative datesets Ciao and Epinions\footnote{Both Ciao and Epinions datasets are available at: http://www.cse.msu.edu/$\sim$tangjili/trust.html} for the Top-K recommendation. As these two datasets provide users' explicit ratings on items, we convert them into 1 as the implicit feedback. This processing method is widely used in previous works on recommendation with implicit feedback~\cite{rendle2009bpr}. We randomly split the user-item interactions of each dataset into training set ($80\%$) to learn the parameters, validation set ($10\%$) to tune hyper-parameters, and testing set ($10\%$) for the final performance comparison~\cite{fan2019Graph}.  The statistics of these two datasets are presented in Table~\ref{tab:dataset}.

In order to evaluate the quality of the recommender systems, we use two popular performance metrics for Top-K recommendation~\cite{wang2017irgan}: Precision@K and Normalized Discounted Cumulative Gain (NDCG@K). We set K as 3, 5, and 10.  Higher values of the Precision@K and NDCG@K indicate better predictive performance.


\begin{table}[tbp]
\centering
\caption{Statistics of the datasets.}
\label{tab:dataset}
\vskip -0.05in
\scalebox{0.85}{
\begin{tabular}{l|c|c}
\hline
Datasets               & Ciao  &Epinions \\ \hline \hline
\# of Users           & 7,317   &14,575 \\ \hline
\# of Items           & 10,4975  &155,527 \\ \hline
\# of Interactions          & 283,319  &418,936\\ \hline
Density of Interactions          & 0.0368\% &0.0184\%  \\ \hline \hline
\# of Social Relations & 111,781 &249,586 \\ \hline
Density of Social Relations  & 0.2087\% &0.1175\% \\ \hline
\end{tabular}}
\end{table}

To evaluate the performance, we compared our proposed model \textbf{DASO} with four groups of representative baselines, including traditional recommender system without social network information (\textbf{BPR}~\cite{rendle2009bpr}), tradition social recommender systems (\textbf{SBPR}~\cite{zhao2014leveraging} and \textbf{SocialMF}~\cite{jamali2010matrix}), deep neural networks based social recommender systems (\textbf{DeepSoR}~\cite{fan2018deep}  and \textbf{GraphRec}~\cite{fan2019Graph}), and adversarial learning based recommender system (\textbf{IRGAN}~\cite{wang2017irgan}). Some of the original baseline implementations (SocialMF, DeepSoR, and GraphRec) are for rating prediction on recommendations. Therefore we adjust their objectives to point-wise prediction with sigmoid cross entropy loss using negative sampling.


We implemented our method with tensorflow. For the  size of representation $d$, we tested the values of $\left \{ 8, 16, 32, 64, 128, 256 \right \}$. The batch size and learning rate were searched in $\left \{ 16, 32, 64, 128, 512,1024 \right \} $ and $\left \{ 0.0005, 0.001, 0.005, 0.01, 0.05, 0.1 \right \}$, respectively. ReLU is set as the activation function.  Moreover, we tested the value of $\lambda$ on $\left \{ 0.5, 1, 10, 50, 100, 200, 500\right \}$.


\subsection{Performance Comparison of Recommender Systems}
Table~\ref{tab:baselines_results} presents the performance of all recommendation methods on two real-world datasets in terms of Precision@K and NDCG@K. We have the following findings:
  
  
  

 \begin{list}{\labelitemi}{\leftmargin=1em}
    \setlength{\topmargin}{0pt}
    \setlength{\itemsep}{0em}
    \setlength{\parskip}{0pt}
    \setlength{\parsep}{0pt}
  \item SBPR and SocialMF outperform BPR. SBPR and SocialMF utilize both user-item interactions and social relations; while BPR only uses the user-item interactions. These improvements show the effectiveness of incorporating social relations for recommender systems.
  
  \item In most cases, the two deep models, DeepSoR and GraphRec, obtain better performance than SBPR and SocialMF, which are modeled with shallow architectures. These improvements reflect the power of deep architectures on the task of recommendations. 
  
  \item IRGAN achieves much better performance than BPR, while both of them utilize the user-item interactions only. IRGAN adopts the adversarial learning to optimize user and item representations; while BPR is a pair-wise ranking framework for Top-K traditional recommender systems. This suggests that adopting adversarial learning can provide more informative negative samples and thus improve the performance of the model. 
  
  \item Our model DASO consistently outperforms all the baselines. Compared with DeepSoR and GraphRec, our model proposes advanced model components to model user representations in both item domain and social domain. In addition, our model harnesses the power of adversarial learning to generate more informative negative samples, which can help learn better user and item representations.
  \end{list}

\begin{figure}[tbp]
\centering
\vskip -0.05in
{\includegraphics[width=0.45\linewidth]{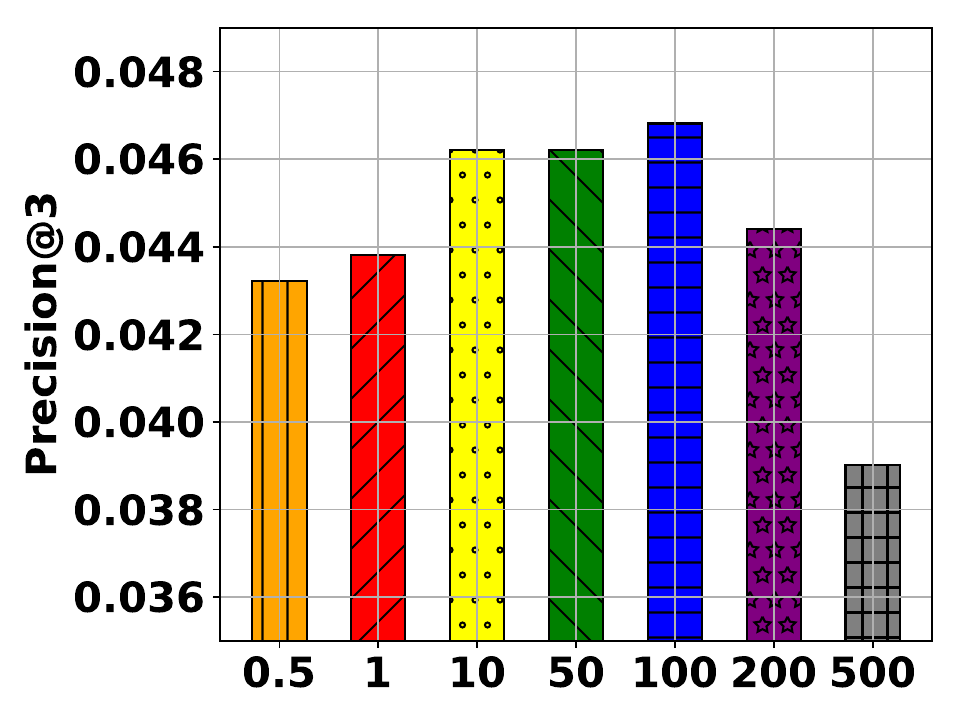}}
\vskip -1ex
\caption{Effect of $\lambda$ on Ciao dataset.}\label{fig:lambda}
 \vskip -0.20in
\end{figure}

\subsubsection{Parameter Analysis}
Next, we investigate how the value of $\lambda$ affects the performance of the proposed framework. The value of $\lambda$ is to control the importance of cycle reconstruction. Figure~\ref{fig:lambda} shows the performance with varied values of $\lambda$ using Precision@3 as the measurement. The performance first increases as the value of $\lambda$ gets larger and then starts to decrease once $\lambda$ goes beyond $100$. 
The performance weakly depends on the parameter controlling the bidirectional influence, which suggests that transferring user’s information from the social domain to the item domain already significantly boosts the performance. However, the user-item interactions and user-user connections are often very sparse, so the bidirectional mapping (Cycle Reconstruction) is proposed to help alleviate this data sparsity problem. Although the performance weakly depends on the bidirectional influence, we still observe that we can learn better user’s representation in both domains.




\section{Related Work}
\label{sec:relatedwork}
As suggested by the social theories~\cite{marsden1993network,wasserman1994social}, people's behaviours tend to be influenced by their social connections and interactions. Many existing social recommendation methods~\cite{fan2018deep,tang2013exploiting,tang2016recommendation,du2017additive,ma2008sorec} have shown that incorporating social relations can enhance the performance of the recommendations. In addition,  deep neural networks have been adopted to enhance social recommender systems.
DLMF~\cite{deng2017deep} utilizes deep auto-encoder to initialize vectors for matrix factorization. DeepSoR~\cite{fan2018deep} utilizes deep neural networks to capture non-linear user representations in social relations and integrate them into probabilistic matrix factorization for prediction. GraphRec~\cite{fan2019Graph} proposes a graph neural networks framework for social recommendation, which aggregates both user-item interactions information and social interaction information when performing prediction.


Some recent works have investigated adversarial learning for recommendation. IRGAN~\cite{wang2017irgan} proposes to unify the discriminative model and generative model with adversarial learning strategy for item recommendation. NMRN-GAN~\cite{wang2018neural} introduces the adversarial learning with negative sampling for streaming recommendation. Despite the compelling success achieved by many works, little attention has been paid to social recommendation with adversarial learning. Therefore, we propose a deep adversarial social recommender system to fill this gap.

\section{Conclusion and Future Work}
\label{sec:conclusion}
In this paper, we present a \textbf{D}eep \textbf{A}dversarial \textbf{SO}cial recommendation model (\textbf{DASO}), which learns separated user representations in item domain and social domain. Particularly, we propose to transfer users' information from social domain to item domain by using a bidirectional mapping method. In addition, we also introduce the adversarial learning to optimize our entire framework by generating informative negative samples. Comprehensive experiments on two real-world datasets show the effectiveness of our model. The calculation of softmax function in item/social domain generator involves all items/users, which is time-consuming and computationally inefficient. Therefore, hierarchical softmax~\cite{morin2005hierarchical,mikolov2013distributed,wang2018graphgan}, which is a replacement for softmax, would be considered to speed up the calculation in both generators in the future direction.

\section{Acknowledgments}
The work described in this paper has been supported, in part, by NSFC-Guangdong Joint Fund under project U1501254, Science Technology and Innovation Committee of Shenzhen Municipality Under project JCYJ20170818095109386, and a start-up fund from the Hong Kong Polytechnic University (project no. 1.9B0V). Tyler Derr, Yao Ma and Jiliang Tang are supported by the National Science Foundation (NSF) under grant numbers IIS-1714741, IIS-1715940, IIS-1845081 and CNS-1815636, and a grant from Criteo Faculty Research Award.

\newpage
\bibliographystyle{named}
\bibliography{references/references}
\end{document}